\documentstyle[preprint,aps]{revtex}

\begin{document}
\draft
\title{\Large On the Chern-Simons Gauge Field}
\author{D.\ G.\  Barci}
\address{Instituto de F\'\i sica, Universidade Federal do Rio de
Janeiro \\ C.P. 68528, Rio de Janeiro, RJ, 21945-970, Brasil}
\author{L.\ E.\ Oxman}
\address{Departamento de F\'\i sica, Facultad de Ciencias Exactas y
Naturales\\
Universidad de Buenos Aires, Ciudad Universitaria, 1428, Buenos
Aires, Argentina}
\date{September 1994}

\maketitle

\begin{abstract}

We show the relationship between a fluid of particles having
charge and magnetic moment in $2+1$ dimensional electromagnetism and
the Chern-Simons statistical field. The matter current which is
minimally coupled to the electromagnetic field has two parts:
the global electromagnetic current, and the corresponding topological
current. The topological current is associated to the induced
electromagnetic current, via a simple constitutive relation between
charges and magnetic moments. We also study the edge states, when the
region that the currents occupy is bounded.

\end{abstract}


\newpage

\section{Introduction}

Over the last years, theories of interacting particles in $2+1$
dimensions have received considerable attention. Perhaps the most
relevant application of these theories is the Quantum Hall Effect
\cite{qhe}.
In this context new phenomena arise connected with the possibility
of statistical transmutation \cite{stat}, or in general with the
possibility of associating charges and fluxes.

Within this approach, theories involving gauge fields with
Chern-Simons dynamics became very popular \cite{wilchek}.
For instance,  the generally accepted explanation of the Fractional
Quantum  Hall Effect (FQHE) is based on the idea that the ground
state is basically the ground state of the Integer Quantum Hall
Effect (IQHE) for composite objects having charge and flux
\cite{jain}.
The more accurate ground state known up to now is the $N$-particle
Laughlin wave function \cite{laughlin}
\begin{equation}
\psi(z_1\ldots z_n)=\prod_{i<j} (z_i -z_j )^m
\prod_i e^{-\frac{1}{4 l^2}|z_i|^2}
\label{gs}
\end{equation}
where $l$ is the magnetic length. Eq. (\ref{gs}) can be associated
with a system of $N$ particles
having an elementary charge $e$ and $m-1$ elementary fluxes, where
$m$ must be an odd integer in order for the wave function to be
antisymmetric.

The implementation of this idea in Quantum Field Theory is performed
through the Chern-Simons theory \cite{fradking}. By coupling the
matter current to a ``statistical'' gauge field $a_\mu$ with
Chern-Simons dynamics
\begin{equation}
S=\int d^3x~J^\mu a_\mu-\frac{\theta}{2}
\epsilon^{\mu\nu\rho}a_\mu\partial_\nu a_\rho ~~~~~,
\end{equation}
the Euler-Lagrange equations lead to the constraints
\begin{equation}
\theta \epsilon^{\mu\nu\rho}\partial_\nu a_\rho = J^\mu
\label{const-cs}
\end{equation}
Note that these equations completely determine the field $a_\mu$ from
the matter configuration. It is clear that these constraints
associate every particle with a ``statistical'' magnetic field and a
``statistical'' electric field perpendicular to the current, both
fields being localized at the position of the particle.

The parameter $\theta$ has $[charge/flux]$ units and can be
expressed as $\theta=\alpha \frac{e}{\phi_0}$, where $e$ is the
electronic charge and $\phi_0$ is the flux quantum.
In other words, $1/\theta$ is the number of quantum statistical
fluxes attached per unit charge.

The complete action including a Maxwell gauge field is
\begin{equation}
\label{maxwell-cs}
S=\int d^3x~{\cal L}_M-\frac{1}{4}F^{\mu\nu}F_{\mu\nu}+J^\mu A_\mu+
J^\mu a_\mu-\frac{\theta}{2} \epsilon^{\mu\nu\rho}a_\mu\partial_\nu
a_\rho~~~,
\label{M+CS}
\end{equation}
where ${\cal L}_M$ is the matter lagrangian.
This action is quadratic in the gauge fields; then, changing to
euclidean space, and by conveniently fixing the gauge we can
integrate out the fields $A_\mu$ and $a_\mu$ to obtain the effective
action for the matter fields
\begin{equation}
\label{effective-cs}
S_{eff}=\int d^3x~{\cal L}_M -\int d^3x~d^3y~\left(
\frac{1}{2}J^\mu\frac{1}{\Box}J_\mu +
\frac{1}{2\theta}\epsilon^{\mu\alpha\beta}J_\mu\partial_\alpha
\frac{1}{\Box}J_\beta \right)
\end{equation}
the third term corresponds to an electromagnetic interaction between
the currents $J$, while the last term comes from the Chern-Simons
sector of the original action and is responsible for the statistical
transmutation of the matter fields.
This mechanism of associating charge and flux is widely accepted in
the  construction of Landau-Ginzburg theories for the
FQHE\cite{lopes-fradking}. Nevertheless, the interpretation
of the ``statistical field'' is not clear.

Recently, it was proposed \cite{marino} that the Chern-Simons term
could be obtained by projecting a topological term from $3+1$ to
$2+1$ dimensions; in this way the statistical field is associated
with the gauge field of a ``modified'' $(3+1)$ QED.
We present in this paper a model which is appropiate for the
description of a fluid of vortices that interact electromagnetically
in $2+1$ dimensions. This description leads to statistical
transmutation and has a very clear interpretation that could allow
future studies of the real dynamics of the charge-flux association.

\section{The Model}

If we are interested in describing the long distance behavior of
particles having a given micro-structure, as a
first approximation, we can consider the first moments of the
internal charge and current distributions. Let us supposse that, in
the frame where the micro-structure is globally at rest, the first
non-zero moments are the total charge and the magnetic dipole moment.
Then, we can assume that we are working with a fluid of
``particle-like'' vortices. In this approximation we can think of
each constituent as having a charge, and an intrinsic magnetic moment
arising from microscopic currents moving on ``circuits'' or
``loops''.
These loops can be taken as a way of representing the intrinsic
currents of each micro-structure, in order to facilitate the
understanding of the origin of the induced currents.

If we consider a neutral loop at rest in a plane, it is easy to see
that there is a magnetic field (in $(2+1)$ electromagnetism) which
is only non-zero inside the loop. This is to be considered as an
approximation, when we look at the local effects, to the real
situation in $(3+1)$ electromagnetism where the magnetic field is not
localized at the loop.
Now, if we consider a neutral loop with constant velocity $\vec{v}$,
it is a simple task to evaluate the corresponding electromagnetic
fields (by Lorentz transforming). The result is that the
electromagnetic field is localized in the inner part of the loop. The
electric field is perpendicular to $\vec{v}$ and comes from
the appearence of a non-zero charge density (having total charge
equal to zero) due to the movement of the loop. Note also that the
velocity $\vec{v}$ is proportional to the ``global'' current
associated to the motion of the loop.

Now, let us consider a ``fluid'' of particle-like vortices. In order
to describe it, we give the global distribution of charge by a
density $J^0_{global}$ and a current density $\vec{J}_{global}$, due
to the global motion of the charged vortices. We also give a
constitutive relation
which simply says that the internal state is the same for all the
particles of the fluid. That is, all the particles have the same
total charge $Q$ and, when considered at rest, they display the same
magnetic dipole moment $\mu \hat{z}$ ($\hat{z}$ is a versor
perpendicular to the plane in consideration). Then, in the case where
the charge density $J^0_{global}$ is stationary and
$\vec{J}_{global}=0$, in every place where we have a charge $Q$ we
also have a magnetic dipole moment $\mu \hat{z}$ and therefore
the magnetic dipole moment density is given by (from now on, we will
simply write $J_{\rho}$ for the global charge and current densities)
\begin{equation}
\label{magnetization}
\vec{m}(\vec{x})=\mu /Q\, J_0 (\vec{x})\hat{z}
\end{equation}
This density implies an induced electric current (due
to the possible lack of compensation of currents coming from
different loops) given by
\begin{equation}
\vec{J}_{ind}=\vec{\nabla}\times\vec{m}
=g\vec{\nabla}\times(J_0\hat{z})=g\hat{z}\times\vec{\nabla}J_0
\label{ind}
\end{equation}
where we have called $g=\mu /Q$.
Written in two-dimensional notation this current takes the form
\begin{equation}
J_{ind}^i=g \epsilon^{0ik}\partial_k J_0
\end{equation}

The electromagnetic interaction lagrangian is obtained by
minimally coupling the total current to the electromagnetic
field,
\begin{equation}
{\cal L}_{int}= A_\mu J^{\mu}_{total}
\end{equation}
Using $J^\mu_{total}=J^\mu+J^\mu_{ind}$ and recalling that
$\vec{J}=0$, $J^0_{ind}=0$ we get
\begin{equation}
{\cal L}_{int}=A_0 J^0 +g A_i\epsilon^{0ik}\partial_k J_0
\end{equation}
The Lorentz-invariant generalization of this interaction lagrangian
is
\begin{eqnarray}
\lefteqn{{\cal L}_{int}=
A_\mu(J^\mu+ g \epsilon^{\mu\nu\rho}\partial_\nu J_\rho)}
\nonumber \\
&&=A_\mu(J^\mu+ gG^\mu)\makebox[.5in]{,}
G^\mu=\epsilon^{\mu\nu\rho}\partial_\nu J_\rho
\label{coup}
\end{eqnarray}
$G^\mu$ is the topological current ($\partial_\mu
G^\mu=0$) associated to the global current $J_\rho$, and gives (up to
the constant $g$) the induced charge density and induced currents in
a general situation.

The interpretation of an interaction term proportional to
$J_\mu \tilde{\cal F}^\mu$, $\tilde{\cal F}^\mu =
\epsilon^{\mu\nu\rho}\partial_\nu A_\rho$ as associated
to particles having magnetic moment has also been given in Refs.
\cite{Kuns} and \cite{stern}. This term coincides with
that given in (\ref{coup}) up to boundary terms, which are relevant
when the region that the currents occupy is bounded (see section IV).
In Ref.\ \cite{Kuns} it is also shown the relationship between this
term and a Chern-Simons term, when the matter is given by a scalar
field that undergoes spontaneous symmetry breaking (SSB). In the next
section we will see that the relation between (\ref{coup}) and the
Chern-Simons theory is more general and does not depend on SSB.

\section{Relationship with the Chern-Simons Theory}

According to (\ref{coup}), the lagrangian for a fluid of
particle-like vortices whose total current (global and induced) is
minimally coupled to the electromagnetic field is
\begin{equation}
{\cal L}=-\frac{1}{4}F_{\mu\nu}F^{\mu\nu}+ A_\mu(J^\mu + gG^\mu)+
{\cal L}_M
\label{els}
\end{equation}
The Euler-Lagrange equation for the electromagnetic field is (in the
gauge $\partial_\mu A^\mu=0$)
\begin{equation}
\Box A^\mu = -(J^\mu+ g G^\mu)
\end{equation}
Given a matter distribution we can write
\begin{equation}
A^\mu = {\cal A}^\mu + b^\mu
\label{Aa}
\end{equation}
where ${\cal A}^{\mu}$ is the electromagnetic field due to the global
currents, and $b^{\mu}$ is the field having sources on the induced
currents:
\begin{equation}
\Box {\cal A}^\mu = -J^\mu \makebox[.5in]{,}
\Box b^\mu = -g G^\mu = -g\epsilon^{\mu\alpha\beta}\partial_\alpha
J_\beta
\end{equation}
Multiplying the second equation by
$\epsilon_{\mu\nu\rho}\partial^\rho$, we obtain
\begin{equation}
\Box \epsilon_{\mu\nu\rho}\partial^\rho b^\mu  = -g
\partial_\nu (\partial .J) +g \Box J_\nu
\end{equation}
using that $J$ is a conserved current, we get the equation
$\Box (\epsilon_{\mu\nu\rho}\partial^\rho b^\mu -g J_\nu)=0$,
and up to a solution to the homogeneous equation of the form
$\epsilon_{\mu\nu\rho}\partial^\rho c^\mu$, $\Box c^\mu=0$,
which can be absorbed as a radiation field in ${\cal A}^\mu$ (cf.
\ref{Aa}): ${\cal A}^\mu \rightarrow {\cal A}^\mu +c^\mu$, we get
\begin{equation}
J^\mu=\frac{1}{g}\epsilon^{\mu\nu\rho}\partial_\nu b_\rho
\label{const}
\end{equation}
or, by fixing the gauge $\partial_\mu b^\mu=0$,
\begin{equation}
b_\mu(x) =-g \epsilon_{\mu\nu\rho} \partial^\nu \int dx'
J^\rho(x') G(x-x')
\end{equation}
where $G(x-x')$ is a Green function for the operator $\Box$ with a
given boundary condition. For instance, the retarded function is
$G_R (x)=1/(2\pi) Q_+^{-1/2}\theta (t)$ where
$Q_+^{-1/2}=(t^2-\vec{x}^2)^{-1/2}$, if $t^2 > \vec{x}^2$ and
$Q_+^{-1/2}=0$ otherwise.

{}From Eq. (\ref{const}) we can see that the field $b_\mu$ and the
global current $J^\mu$ satisfy a Chern-Simons relation.
Note that in this case $b_\mu$ is a true electromagnetic field and
not the statistical field used in (\ref{const-cs}). The physical
interpretation for the expression (\ref{const}) can be traced back
from the discussion at the beginning of section II: the magnetic and
electric fields having sources in the intrinsic currents of the
moving loops are localized at these loops, the electric field being
perpendicular to the global current.

In Eq. (\ref{ind}), we have introduced the coupling $g=\mu /Q$.  From
Eq. (\ref{Aa}) and (\ref{const}), we find that
$g$ also represents the number of (intrinsic) real flux
quanta per unit charge: $g=\Phi /Q$. Note that if we take a
static vortex at $\vec{x}_0$, the magnetic field is equal to the
magnetic dipole moment density:
$\vec{B}=\vec{m}=\mu \delta (\vec{x}-\vec{x}_0) \hat{z}$, and
integrating over an area containing $\vec{x}_0$ we obtain $\Phi=\mu$,
in accordance with our previous definition $g=\mu /Q$.

Now, we will study the possibility of considering $b$
as a background field. For instance, if we take an external classical
source $J$ (with zero divergence), we can obtain, from the partition
function $Z$, the probability ${\cal P}=Z Z^{\ast}$ to remain in the
vacuum state (vacuum persistence). From (\ref{els}), path integrating
over $A$, and using Feynman's prescription to invert $\Box$ we get
\begin{eqnarray}
Z&=&\exp \left( \frac{i}{2} \int dx dy H_{\mu}(x)\int
\frac{dk}{(2\pi)^3}e^{ik(x-y)} \frac{\eta^{\mu\nu}}{k^2+i\epsilon}
H_{\nu}(y)\right)
\nonumber \\
&=&\exp \left( \frac{i}{2}\int \frac{dk}{(2\pi)^3}
\frac{\tilde{H}_{\mu}(k)\tilde{H}^{\mu}(-k)}{k^2+i\epsilon}\right)
\end{eqnarray}
where $H_\mu = J_\mu +g\epsilon_{\mu \nu \rho}
\partial^{\nu}J^{\rho}$ and $\tilde{H}_{\mu}$ is its Fourier
transform:
\begin{eqnarray}
\tilde{H}_\mu(k) = \tilde{J}_\mu(k) +
g\epsilon_{\mu \nu \rho} k^{\nu}\tilde{J}^{\rho}(k)\nonumber \\
\tilde{H}_\mu(-k) = \tilde{J}^{\ast}_\mu(k) -
g\epsilon_{\mu \nu \rho} k^{\nu}\tilde{J}^{\rho \ast}(k)
\label{TF}
\end{eqnarray}
($J(x)$ is real and therefore we have
$\tilde{J}(-k)=\tilde{J}^{\ast}(k)$).
Now, the probability to remain in the vacuum comes from the real part
($R$) of the exponent in $Z$:
\begin{equation}
{\cal R}=\frac{\pi}{2}\int \frac{dk}{(2\pi)^3} \delta (k^2)
\tilde{H}_{\mu}(k)\tilde{H}^{\mu}(-k)
\makebox[.5in]{,}
{\cal P}=e^{2{\cal R}}
\label{real}
\end{equation}
and using (\ref{TF}) in (\ref{real}), we get
\begin{eqnarray}
\lefteqn{{\cal R}=\frac{\pi}{2}\int \frac{dk}{(2\pi)^3} \delta (k^2)
\left( \tilde{J}_{\mu}(k)\tilde{J}^{\mu \ast}(k) +
g\epsilon_{\mu \nu \rho}
k^{\nu}\tilde{J}^{\rho}(k)\tilde{J}^{\mu \ast}(k)\right.}
\nonumber \\
&&\left. -g\tilde{J}^{\mu}(k)\epsilon_{\mu\nu'\rho'}
k^{\nu'}\tilde{J}^{\rho'\ast}(k)-
g^2 (k^2\tilde{J}(k) .\tilde{J}^{\ast}(k)-
k.\tilde{J}(k)\, k .\tilde{J}^{\ast}(k)\right)\nonumber \\
&&=\frac{\pi}{2}\int \frac{dk}{(2\pi)^3} \delta (k^2)
\tilde{J}_{\mu}(k)\tilde{J}^{\mu \ast}(k)
\label{nog}
\end{eqnarray}
where we have used that $J$ is conserved (note that $k.\tilde{J}=0$,
$k^2=0$ imply ${\cal R}< 0$).
Then, the property of having a non-zero $g$ does not modify the
vacuum survival probability.

The canonically quantized field $A$ can be written as
$A(x)={\cal A}(x)+b(x)$,
$~{\cal A}(x)=\hat{A}(x)-\int dx' G(x-x')J(x')$, where ${\cal A}$
contains the fluctuating part
$\hat{A}$, $\Box \hat{A}=0$. Now, suppose that the field $A$ were
only coupled to the induced current: $Q$ tends to zero while
keeping $\mu =g Q$ finite, that is, $J^\mu \rightarrow 0$ keeping
$gG^\mu$ finite. In this case, from (\ref{nog}), the probability to
remain in the vacuum is equal to $1$; then, from unitarity, the
probability to go from the vacuum to any other state $|n\rangle$
($\langle 0|n\rangle$=0) is equal to zero. Therefore, in this case, a
situation where the field is just the classical part $b$ and the free
fluctuating part $\hat{A}$ is always in the vacuum state is
self-consistent at the quantum level. This differs from the case were
the $A$ field is only coupled to $J$. In this case, according to
(\ref{nog}), the classical current excites particles from the vacuum
state and the field $\hat{A}$ can not be left in the vacuum.

Then, we could expect, in the general case, to be able to replace the
full electromagnetic field by a background field $b$
and a fluctuating field ${\cal A}$ having sources on the global
current $J$. In order to look for this possibility we will suppose
that the current $J$ comes from a (quantum) matter sector
represented by a field $\psi$  and we will consider the
corresponding partition function, obtained from the path integral
\begin{eqnarray}
\lefteqn{Z[H,K]=\int [{\cal D}\psi] [{\cal D}A]
\exp i\int dx (-\frac{1}{4}F^2 +}\nonumber \\
&& +A (J (\psi)+g G (\psi))
+{\cal L}_M (\psi)+{\cal L}_{GF}(A)+HA+K\psi )
\label{partition}
\end{eqnarray}
where ${\cal L}_{GF}=-1/2 (\partial .A)^2$ is the (Lorentz) gauge
fixing term.
In the path integral over $A$, we can make the change of variables
(having jacobian equal to $1$):
\begin{equation}
A^\mu={\cal A}^\mu + b^\mu (\psi)
\end{equation}
where $b^\mu (\psi)=-g \epsilon_{\mu\nu\rho} \partial^\nu \int dx'
J^\rho(\psi(x')) G(x-x')$. If we call ${\cal F}$ and $f$ the
field-strength tensors for ${\cal A}$ and $b$, respectively, the
exponent in (\ref{partition}) now reads
\begin{equation}
-\frac{1}{4}{\cal F}^2 -\frac{1}{2}{\cal F} f -\frac{1}{4}f^2+
({\cal A}+b)J+g {\cal A}G+gbG+{\cal L}_M+
{\cal L}_{GF}({\cal A})+H({\cal A}+b)+K\psi
\end{equation}
Here, ${\cal L}_{GF}({\cal A})=-1/2 (\partial .{\cal A})^2$
($\partial .b=0$). From (\ref{const}) we have
$f_{\mu\nu}= g\epsilon_{\mu \nu \rho}J^\rho$, then (up to
divergences) we obtain
\begin{equation}
-\frac{1}{2}{\cal F}f=-g{\cal A}G~~~,~~~
-\frac{1}{4}f^2=-\frac{g^2}{2} J^2~~~,~~~
gbG=g^2 J^2
\end{equation}
and results
\begin{eqnarray}
\lefteqn{Z[H,K]=\int [{\cal D}\psi] [{\cal DA}]\exp i\int dx
(-\frac{1}{4}{\cal F}^2+({\cal A}+b)J+}\nonumber \\
&&+\frac{g^2}{2} J^2+{\cal
L}_M+{\cal L}_{GF}({\cal A})+H({\cal A}+b)+K\psi )
\label{back}
\end{eqnarray}
Then we find, by taking functional derivatives with respect to $H$
and $K$ that we can compute $n$-point Green functions for
the fields $A$ and $\psi$ using for this
fields the dynamics (\ref{els}) or, equivalently, we can replace $A$
by ${\cal A}+b$ and compute the corresponding Green functions using
the dynamics given by
\begin{equation}
\tilde{{\cal L}}=-\frac{1}{4}{\cal F}^2+{\cal A}J+
{\cal L}_M+ bJ+\frac{g^2}{2} J^2
\label{monio}
\end{equation}
Here, we see that the $b$ field appears as a background and the
${\cal A}$ field has sources only on $J$; the matter lagrangian is
also modified by the presence of a local current-current term.
The $bJ$ term leads to a non-local interaction
\begin{equation}
bJ=-g\int d^3x~d^3y~ \epsilon^{\mu\alpha\beta}J_\mu(x)\partial_\alpha
\frac{1}{\Box}J_\beta(y)\makebox[.5in]{,}J=J(\psi)
\end{equation}
which can be made local by introducing a Chern-Simons field $a$,
with $\theta =1/(2g)$ (cf. (\ref{maxwell-cs}) and
(\ref{effective-cs})):
\begin{equation}
\tilde{{\cal L}}\rightarrow
-\frac{1}{4}{\cal F}^2+{\cal A}J+aJ-\frac{1}{4g}
\epsilon^{\mu\nu\rho}a_\mu\partial_\nu a_\rho + {\cal L}_M +
\frac{g^2}{2} J^2
\label{monio'}
\end{equation}
Then, we can compute a probability amplitud between two
$A$, $\psi$ configurations, using ${\cal L}$ (Eq. (\ref{els})), or we
can compute the corresponding expression for the ${\cal A}$, $\psi$
fields, using $\tilde{\cal L}$ (Eq. (\ref{monio}) or (\ref{monio'})).
Clearly, this result also holds when the current comes from
the motion of a given number of point particles, replacing in the
path integral $\psi$ by the particle's coordinates.

The additional terms in the matter lagrangian can also be understood
if we consider the effective dynamics for the vortices that interact
via $(2+1)$ electromagnetism, obtained by averaging over the gauge
field. From (\ref{els}), integrating out the field $A_\mu$, we
obtain the effective action
\begin{eqnarray}
\lefteqn {S_{eff}= \int d^3x~{\cal L}_M}-\nonumber \\
&& -\frac{1}{2} \int d^3x~ d^3y~(J_\mu(x) + g G_\mu(x))
\frac{1}{\Box} (J^\mu(y) + g G^\mu(y)) \nonumber \\
&&=\int d^3x~{\cal L}_M + \frac{g^2}{2} J_\mu J^\mu -\int d^3x~d^3y~
\left( \frac{1}{2}J^\mu\frac{1}{\Box}J_\mu
+g\epsilon^{\mu\alpha\beta}J_\mu\partial_\alpha
\frac{1}{\Box}J_\beta \right)
\label{effective-topological}
\end{eqnarray}
where $1/ \Box$ stands for $G_F (x-y)$, the Feynman's Green function:
$G_F (x)=1/(2\pi)(Q-i\epsilon)^{-1/2}$, $Q=t^2-\vec{x}^2$.
We have just seen in (\ref{nog}) that the $g$-dependent part of
(\ref{effective-topological}) is real: the second current-current
interaction term in (\ref{effective-topological}) comes from the
interaction of the induced currents via $1/ \Box$,
while the last term comes from the interaction of the induced
and global currents and is known as the Hopf term, which changes the
particle's statistics (for a review, see Ref.\ \cite{Forte}). This
effective action can also be obtained by averaging over ${\cal A}$ in
(\ref{monio}), that is, integrating the ${\cal A}$ field in
(\ref{back}) and setting to zero the external currents.

The current-current term is a contact term, when the matter comes
from the motion of a given number of point particles. In this case,
if we look at the long distance properties of the theory, we can drop
out the current-current interaction in (\ref{monio'}), obtaining the
Maxwell and Chern-Simons model of Eq. (\ref{M+CS}) (making the
identification $\theta =1/(2g)$).

The short distance behavior differs by the presence of a
current-current term. This term distinguishes the interaction of
particles having attached a real electromagnetic flux from the
interaction of particles having attached a statistical flux. This
type of ``velocity dependent contact interaction'' has been
considered in Ref.\ \cite{stern}, in the context of classical
Maxwell-Chern-Simons theory.


\section{Edge States}

It is well known that the Chern-Simons action is gauge invariant up
to boundary terms. So if we work in a bounded region, in order to
make the theory gauge invariant, it is necessary to add a $1+1$
dimensional action to the $2+1$ dimensional action
(\ref{maxwell-cs}).
In the case of the Quantum Hall Effect this will produce edge
currents that satisfy a Kac-Moody algebra \cite{wen1}.
These edge currents appear to be fundamental when explaning the
universal aspects of the QHE \cite{wen2}\cite{stone}.

In our case, to maintain the gauge invariance of the theory it is
sufficient to require the local conservation of the total electric
charge. The induced part of the total current is automatically
divergenceless, as it corresponds to a topological current. This
can be related to the fact that the induced current comes from the
part of the vortex charge distribution that is a magnetic moment in
the vortex rest-frame, and gives no net charge transportation.
Then, the charge is locally conserved when the global current is
divergenceless everywhere. Note that when the currents vanish outside
a finite region $D$, we can write
\begin{equation}
J_\rho (t,\vec{x}) = j_\rho (t,\vec{x})\, \theta (D)
\label{Jota}
\end{equation}
where $\theta (D)$ is a function whose value is one when $\vec{x}$ is
inside $D$ and zero otherwise. If the boundary is static we have
\[
\partial^{\rho} J_\rho (t,\vec{x}) = \partial^{\rho} j_\rho
(t,\vec{x})\, \theta (D) + j_i (t,\vec{x})\, \partial^i \theta (D)
\]
then, in this case, the charge will be locally conserved (even at the
edge) if $j_\rho$ is divergenceless in the bulk, and the spatial part
of $j_\rho$ is tangential to the border of the bounded region.
Under these circumstances, the theory will be gauge invariant.

{}From (\ref{Jota}) we see that the induced current has two terms (cf.
(\ref{coup}))
\begin{equation}
gG^\mu =g\theta (D)\, \epsilon^{\mu\nu\rho}\partial_\nu j_\rho +
g\epsilon^{\mu\nu\rho}j_\rho \partial_\nu \theta (D)
\label{bb}
\end{equation}
the first term in (\ref{bb}) has support at $D$ and gives the induced
currents in the bulk: $gG_{bulk}$, while the second term gives the
induced currents at $\partial D$ (the border of
$D$): $gG_{\partial D}$. If the border is static, $\theta (D)$ is
time-independent and we obtain for the second term
\begin{equation}
-g\epsilon^{\mu k\rho}j_\rho n_k \delta (\partial D)
\end{equation}
where $n_k$ is the $k$-th component of the versor contained in the
plane, which is normal to $\partial D$ (and external to $D$), and
$\delta (\partial D)$ is a delta-function with support at $\partial
D$. Then, the charge and current density at $\partial D$ are
\begin{equation}
gG^0_{\partial D}=-g\epsilon^{ki}j_i n_k \delta (\partial D)=
-g\, \vec{j}.\hat{t}\, \delta (\partial D)
\makebox[.5in]{,}
g\vec{G}_{\partial D}=gj_0\, \hat{t}\, \delta (\partial D)
\label{border}
\end{equation}
where $\hat{t}=\hat{z}\times\hat{n}$ is the tangent versor to
$\partial D$. The physical interpretation of the charge and current
densities at the edge is clear: the charge density is due to the fact
that the moving loops imply a distribution of electric dipole
moment, perpendicular to the current, and the associated charge
distribution is not compensated at the edge. Similarly, the induced
current comes from the lack of compensation of the intrinsic currents
at the edge.

If a part of $D$ is given by the points bellow $y=0$, then the edge
densities at $y=0$ are
\begin{equation}
\label{y=0}
gG^0_{\partial D}=+g j_1 \delta (y)
\makebox[.5in]{,}
gG^1_{\partial D}=-g j_0 \delta (y)
\end{equation}
this can be written using two-dimensional notation as
\begin{equation}
gG^{\mu}_{\partial D}=g\epsilon^{\mu \nu} j_\nu \delta (y)
\makebox[.5in]{,} \mu=0,1
\end{equation}
If the border of $D$ changes with time, as it occurs for a
droplet of fluid, we should consider in (\ref{bb}) the time
derivative of $\theta (D)$; note however that the expression for the
charge density at the border does not depend on this derivative and
is still given by
\begin{equation}
gG^0_{\partial D}=-g\, \vec{j}.\hat{t}\, \delta (\partial D)
\label{border'}
\end{equation}
where $\partial D$ is the border of the droplet, which can change
with time.

Recalling that the time-component of the topological current
represents the induced charge density coming from the (intrinsic)
magnetic moment part of each micro-structure, we have that the total
induced charge (in the bulk and at the edge) must be zero. This can
be verified as the induced charge in the bulk is (cf. \ref{bb})
\begin{equation}
g\int_D d^2x~ \epsilon^{ik}\partial_i j_k =
g\oint_{\partial D} \vec{j}. d\vec{\ell} =
g\oint_{\partial D} \vec{j}.\hat{t}\, dl
\end{equation}
which cancels against the induced charge at $\partial D$ (cf.
\ref{border'}):
\begin{equation}
Q_{\partial D}= -g\oint_{\partial D} \vec{j}.\hat{t}\, dl
\label{tic}
\end{equation}
We can see that the $1+1$ quiral currents that are necessary to
restore the gauge invariance of the Chern-Simons theory are the
analog of the induced charge and current densities, at the edge of a
fluid of particle-like vortices. Now, we may ask about the conditions
that lead to the conservation of the induced charge at $\partial D$.
The quantum states carrying this charge can be constructed from an
operator $L$ that satisfies
\begin{equation}
Q_{\partial D} L|0\rangle= q L|0\rangle
\end{equation}
In other words, the equation
\begin{equation}
\label{Leq}
[Q_{\partial D},L]= q L
\end{equation}
must be solved.

In the case of a linear infinite edge ($y=0$), the currents are given
by (\ref{y=0}) and the topological charge can be written as
\begin{equation}
Q_{\partial D}=g \int_{-\infty}^{\infty} j(x)~dx
\end{equation}
where $j=j_1(x)$.
It is well known \cite{wen1}\cite{wen2}\cite{stone} that if the
dynamics of the model leads to an incompressible fluid state, then
the edge currents will satisfy a Kac-Moody algebra
\begin{equation}
\label{algebra}
[j(x),j(x')]=i\frac{d}{dx}\delta(x-x')
\end{equation}
In this case, the operators that obey (\ref{Leq}) can be
characterized by:
\begin{equation}
L_{\alpha}= exp\left\{-i\int_{-\infty}^{\infty} \alpha(x)
j(x)~dx\right\}
\label{ripplon}
\end{equation}
where $\alpha(x)$ is an arbitrary function.
Using this expression and (\ref{algebra}), we immediatly obtain
the induced charge at the border:
\begin{equation}
q=g[\alpha(+\infty)-\alpha(-\infty)]
\label{ctopo}
\end{equation}
Similarly, in the case of a droplet of incompressible fluid, the
states are characterized by a function $\alpha (\theta)$ (where
$\theta$ parametrizes the border) and the induced charge is $g[\alpha
(2\pi)-\alpha (0)]$. That is, the topological charge
$q$ that a state $|\alpha\rangle=L_\alpha |0\rangle$ supports
only depends on the boundary conditions on $\alpha$.

These excitations $|\alpha\rangle$ that occur at the border of an
incompressible quantum fluid are the so called ``ripplons'' and have
been introduced in Ref.\ \cite{stone}. These ``ripplons'' were shown
to correspond to deformations of the droplet surface that move along
the surface with a fixed velocity and their shape unchanged. For
instance, for a linear edge, the ripplon evolves according to
$\alpha (x)\rightarrow \alpha(x-vt)$ and the charge $q$,
given by (\ref{ctopo}), does not change. Then, in general, we see
that in the case of an incompressible fluid of particle-like
vortices, the induced electric charge at the border is conserved.

\section{Discussion}

In this work we have considered a fluid of particle-like vortices
interacting electromagnetically in $2+1$ dimensions. The matter
current has two parts: a global current associated to the motion
of the total charge of each vortex, and an induced current arising
from the lack of compensation of the intrinsic currents coming from
different vortices. By considering a simple constitutive relation
(which says that the internal state of every vortex is the same), we
can write the induced current as the topological current associated
to the global current. Then, given a matter distribution we obtain a
Chern-Simons relation between the global current and the
electromagnetic field having sources on the induced currents.

At the quantum level, we have found that the n-point Green functions
for a system where the electromagnetic field is minimally coupled
to the global current $J$ and the induced (topological) current
$gG$ (Eq. (\ref{els})) can be computed from a theory containing a
Maxwell field and a Chern-Simons field (with $\theta=1/(2g)$)
minimally coupled to $J$, plus a local current-current term that
modifies the matter lagrangian (Eq. (\ref{monio'})).
In particular, after averaging over the gauge fields, the effective
dynamics for both models is the same, except for the
current-current term coming from the interaction of the
intrinsic electromagnetic fields of a given vortex with the intrinsic
currents of the other.
If the matter current comes from the motion of point particles, the
current-current term is a contact term which is unimportant, when
we consider the long distance behavior of the model. In this case we
make contact with Maxwell's theory with a Chern-Simons field.

Our model is constructed just in terms of elecromagnetic fields;
then, unlike the Chern-Simons construction, it is gauge invariant by
itself even when the region that the currents occupy is bounded.
Note that there are induced currents localized at the edge
which are the analog of the quiral currents needed to restore the
gauge symmetry in the Chern-Simons theory. These currents can be used
to characterize the edge states, as the correponding induced charge
is conserved for an incompressible quantum droplet.

In this article we have not been concerned with the possible
mechanisms that can lead to the dynamical formation of structures
having charge and flux. It may be possible \cite{daniel} that in
non-relativistic systems the competition between the
electron-electron interaction and the interaction with a strong
external magnetic field could produce
effective objects that correspond to the quiral structures we have
considered here.
In this framework, the understanding of the FQHE would be related
to the understanding of the origin of micro-structures with a
given relation of $[flux/charge]$ per structure. Given this
particle-like vortices, their effective dynamics could be described
by the lagrangian (\ref{monio'}), and it would be interesting to
explore the system's response to external electromagnetic fields in
order to compute physical parameters as the conductivity.

\section*{acknowledgments}
We would like to thank Prof. C.\ A.\ A.\ de Carvalho,
Prof.\ E.\ C.\ Marino and R.\ Cavalcanti for usefull discussions on
related matters.

This work was partially supported by CNPq, Brasil, and CONICET,
Argentina.


\newpage
\begin{thebibliography}{99}
\bibitem{qhe} ``The Quantum Hall Effect'',  R.\ E.\ Prange and S.\
M.\ Girvin (Springer-Verlag, 1987)
\bibitem{stat} ``Fractional Statistics and Anyon Superconductivity'',
F.\ Wilczek (World Scientific, 1990)
\bibitem{wilchek} F.\ Wilczek, Phys.\ Rev.\ Lett {\bf 48}, 1144
(1982).
\bibitem{jain} J.\ Jain, Phys.\ Rev.\ {\bf B40}, 8079 (1989)
\bibitem{laughlin} R.\ B.\ Laughlin, Phys.\ Rev.\ Lett.\ {\bf 50},
1395 (1983)
\bibitem{fradking} Eduardo Fradkin, Proceedings of Physics at High
Magnetic Fields, FSU, Thallahassee, FL, May 14-18, 1991.
\bibitem{lopes-fradking} Ana Lopez and Eduardo Fradkin, Phys.\ Rev.\
{\bf B 44}, 5246 (1991).
\bibitem{marino} E.\ C.\ Marino, Nuc.\ Phys.\ {\bf B 408} [FS] 551,
(1993).
\bibitem{Kuns} M.\ E.\ Carrington and G.\ Kunstatter, University of
Winnipeg Preprint WIN-94-04
\bibitem{stern} J.\ Stern, Phys.\ Letts.\ {\bf B265}, 119 (1991)
\bibitem{Forte} S.\ Forte, Rev.\ Mod.\ Phys.\ {\bf 64}, No. 1, 193
(1992)
\bibitem{wen1} X.\ G.\ Wen, Phys.\ Rev.\ {\bf B 43}, 11025 (1991).
\bibitem{wen2} X.\ G.\ Wen, Int.\ J.\ Mod.\ Phys.\ {\bf B 6}, 1711
(1992)
\bibitem{stone} M.\ Stone, Ann.\ Phys.\ (N.Y.) {\bf 207}, 38 (1991)
\bibitem{daniel}C.\ Arag\~ao de Carvalho,  D.\ G.\ Barci and L.\
Moriconi, Phys.\ Rev.\ {\bf  B50}, 4648 (1994)
\end{thebibliography}
\end{document}